\documentclass[a4paper,fleqn,usenatbib]{mnras}
\usepackage{newtxtext, newtxmath, graphicx,enumerate,amsmath, morefloats,amstext, amssymb, amsmath}
\usepackage[T1]{fontenc}
\usepackage{ae,aecompl}
\usepackage[figure,figure*]{hypcap}

\newcommand{\oiiw}{\mbox{[\ion{O}{ii}] $\lambda\lambda$3726,3729}}

\newcommand{\niibw}{\mbox{[\ion{N}{ii}] $\lambda$6583}}
\newcommand{\niiw}{\mbox{[\ion{N}{ii}] $\lambda \lambda$6548,6583}}
\newcommand{\niitw}{\mbox{[\ion{N}{ii}] $\lambda$5755}}

\newcommand{\siiw}{\mbox{[\ion{S}{ii}] $\lambda \lambda$6716,6731}}

\newcommand{\oiiibw}{\mbox{[\ion{O}{iii}] $\lambda$5007}}

\newcommand{\oiiitw}{\mbox{[\ion{O}{iii}] $\lambda$4363}}

\newcommand{\siitw}{\mbox{[\ion{S}{ii}] $\lambda$4068,4076}}
\newcommand{\oiitw}{\mbox{[\ion{O}{ii}] $\lambda$7320,7330}}
\newcommand{\caiiaw}{\mbox{[\ion{Ca}{II}] $\lambda$7291}}
\newcommand{\caiibw}{\mbox{[\ion{Ca}{II}] $\lambda$7324}}

\newcommand{\oii}{\mbox{[\ion{O}{ii}]}}
\newcommand{\nii}{\mbox{[\ion{N}{ii}]}}
\newcommand{\hal}{\mbox{H$\alpha$}}
\newcommand{\sii}{\mbox{[\ion{S}{ii}]}}
\newcommand{\hb}{\text{H$\beta$}}

\title[N/O in Quiescent Galaxies]{Nitrogen-to-Oxygen Abundance Ratio Variation In Quiescent Galaxies}
\author[R. Yan]{Renbin Yan$^{1}$\thanks{E-mail: yanrenbin@uky.edu}
\\
$^{1}$Department of Physics and Astronomy, University of Kentucky, 505 Rose Street, Lexignton, KY, 40506; yanrenbin@uky.edu}

\date{Accepted XXX. Received YYY; in original form ZZZ}
\pubyear{2017}

\begin{document}
\label{firstpage}
\pagerange{\pageref{firstpage}--\pageref{lastpage}}
\maketitle

\begin{abstract}
For the first time, we establish a gas phase abundance pattern calibration for quiescent galaxies using optical emission lines. Quiescent galaxies have warm ionized gas showing line ratios similar to low-ionization nuclear emission line regions (LINER). The ionization mechanism for the gas is still an unsettled puzzle. Despite the uncertainty in the ionization mechanism, we argue that we can still infer certain gas phase abundance pattern from first principles. We show that the relative trend in N/O abundance can still be reliably measured based on \niiw/\oiiw\ and a direct measurement of the electron temperature. We construct a composite direct temperature tracer that is independent of extinction correction, by combining \oiiw/\oiitw\ and \siiw/\siitw\ and canceling out the effect of extinction, as these involve the easiest-to-detect auroral lines in quiescent galaxies. With theoretical modeling, we establish the calibration for N/O based on \nii/\oii\ and a temperature tracer. We apply this technique to quiescent galaxies in the nearby Universe and find they span a range of 0.35 dex in N/O ratio from 17-percentile to 83-percentiles of the whole distribution. These measurements can shed light on the chemical enrichment history of the warm ionized gas in quiescent galaxies.


\end{abstract}



\section{Introduction}
Galaxies that are not forming stars can be referred to as quiescent galaxies. In the past, quiescent galaxies are often thought to contain no neutral gas. But this view has been challenged by recent observations of molecular and atomic gas in these galaxies\citep[e.g.][]{DavisT11,Serra12,Young14}. Optical spectroscopy observations also revealed weak optical emission lines in these galaxies indicating warm ionized gas with a temperature around $10^4$K and dust\citep{Phillips86,Kim89,Buson93,Goudfrooij94,Macchetto96,Zeilinger96,Lauer05,Sarzi06,DavisT11, Singh13, Belfiore16, Gomes16}. The relative intensity ratios among the optical emission lines indicate their production must be different from the photoionization by young hot stars in star-forming galaxies. However, we do not yet know for sure what physical mechanism produced these warm ionized gas. Despite this uncertainty, there can be abundance pattern variations we can measure based on direct temperature measurements. Because the abundance pattern of the interstellar medium provides the fossil record of the chemical enrichment history, measuring the elemental abundance of the gas will provide important  insights into the star formation history of these galaxies and the baryonic flows in and out of them. 


Gas phase abundances can be estimated using many different wavebands. Using X-ray observations, one could measure the iron abundance in the hot gas in massive ellipticals and galaxy clusters \citep[e.g.][]{EdgeS91, AllenF98}. However, the lighter elements are much more difficult to measure in X-rays. The lighter elements are much more easily detectable with optical emission lines, which probe the warm ionized gas component. With optical emission lines, gas metallicity are usually measured only for star-forming galaxies as we understand well the ionization mechanism in star-forming regions. For quiescent galaxies, there have been very few attempts before. \cite{AtheyB09} applied the R23 calibration derived from star-forming region models on early-type galaxies. They assumed that the gas is photoionized by post-Asymptotic Giant Branch stars (post-AGB stars) and neglected the difference in the ionizing spectra between post-AGB stars and that of massive OB stars. This can lead to very large systematic uncertainty in the resulting oxygen abundance, and can also change the relative abundance difference between galaxies. \cite{Storchi-Bergmann98} did similar analysis for LINERs but that study suffers from the same issue. 

The ionization mechanism for the warm ionized gas in quiescent galaxies has been hotly debated for more than 20 years. Several ionization mechanisms have been proposed to explain the line ratios observed in them. These include photoionization by an active galactic nucleus \citep{FerlandN83,HalpernS83,GrovesDS04II}, photoionization by post-AGB stars \citep{Binette94, Stasinska08}, photoionization by the hot X-ray emitting gas \citep{VoitD90, DonahueV91}, collisional ionization by fast shocks \citep{DopitaS95}, and heat exchange through conduction or turbulent mixing layers between hot and cold gas \citep{SparksMG89,Slavin93}. Among these, photoionization by AGN has been largely ruled out as the dominant mechanism in the great majority of quiescent galaxies \citep{Sarzi10,YanB12,Singh13,Belfiore16,Gomes16}. However, the jury is still out on which of the other mechanisms is the dominant source. Photoionization by post-AGB stars is often considered to be the most likely source, but this is not confirmed. A recent investigation by \cite{Yan18a} found that neither photoionization nor shocks could explain all the line ratios measured in the stacked spectra of these quiescent galaxies.

Despite this uncertainty in the ionization mechanism, we argue in this paper that it is still possible to measure the abundance ratios among a few elements. For example, we can measure N/O abundance ratio as long as we can get a measurement of \niiw/\oiiw\ and a measurement of electron temperature. This is because, regardless of the ionization mechanism, the \niiw\ and \oiiw\ lines are always collisionally-excited and the ratio between them only depends on N$^+$/O$^+$ and the temperature. Nitrogen and Oxygen also have very similar structures in their ionization potentials: the energies to ionize neutral N and O to N$^+$ and O$^+$ differ by only $0.9$eV, while the energies to ionize these to the next level differ by $5.5$eV. This means that, as long as the energy distribution of the ionizing photons or particles is reasonably smooth, Nitrogen and Oxygen would have nearly the same fraction of atoms in a singly ionized state and those ions would be found in the same spatial region, regardless of the ionization mechanism. This assumption should hold for all the ionization mechanisms mentioned above. Therefore, we can establish a calibration for N/O abundance ratio in quiescent galaxies, although we have not fully settled the ionization mechanism for the gas.

We use photoionization models to derive the calibration, while knowing the calibration is not just limited to photoionization models. Different ionization mechanism may cause small systematic shifts in the absolute N/O abundance ratios due to inaccuracies in the above assumptions. But the relative trend should remain robust. We then apply this to measure the N/O abundance in quiescent galaxies in the nearby Universe. 

Our approach is similar to that of the metallicity calibrations for star-forming galaxies, but we improve on it in three important aspects. First, we do not assume a fixed relationship between N/O and O/H in our theoretical modeling. Instead, we establish a grid of models sampling a wide range of N/O vs. O/H patterns. Second, we construct a direct electron temperature tracer that is independent of extinction correction. This is based on two direct temperature tracers which have been rarely used due to their sensitivity to extinction. We combine them together to cancel out the extinction effect and form an extinction-insensitive temperature tracer. Third, the calibration we derive will link line ratio measurements directly with abundance ratios, without going through an intermediate step of deriving temperatures and ionization correction factors. The latter has been the approach adopted in many $T_e$-based abundance calibrations. Our approach will reduce the number of necessary assumptions and the systematic errors associated with those assumptions.

In this paper, we first describe the setup of the simulations(\S\ref{sec:simulation}), then establish the N/O abundance calibration(\S\ref{sec:calibration}), and finally apply the calibration on SDSS galaxies to measure their N/O abundance ratio (\S\ref{sec:data}).

\section{Simulations}\label{sec:simulation}

We run a large number of Cloudy simulations to sample the potential parameter space. We use version 17.00 of Cloudy, which is last described by \cite{Ferland17}. There are many physical parameters that can affect the resulting line ratios. The abundance pattern, the depletion of elements onto dust, the ionization mechanism, the density or pressure of the interstellar medium. Here, as we are most concerned with the abundance pattern, we try to reduce the number of assumptions we make about it. We run the simulation for as many different abundance patterns as possible. 

\begin{figure}
\begin{center}
\includegraphics[width=0.48\textwidth]{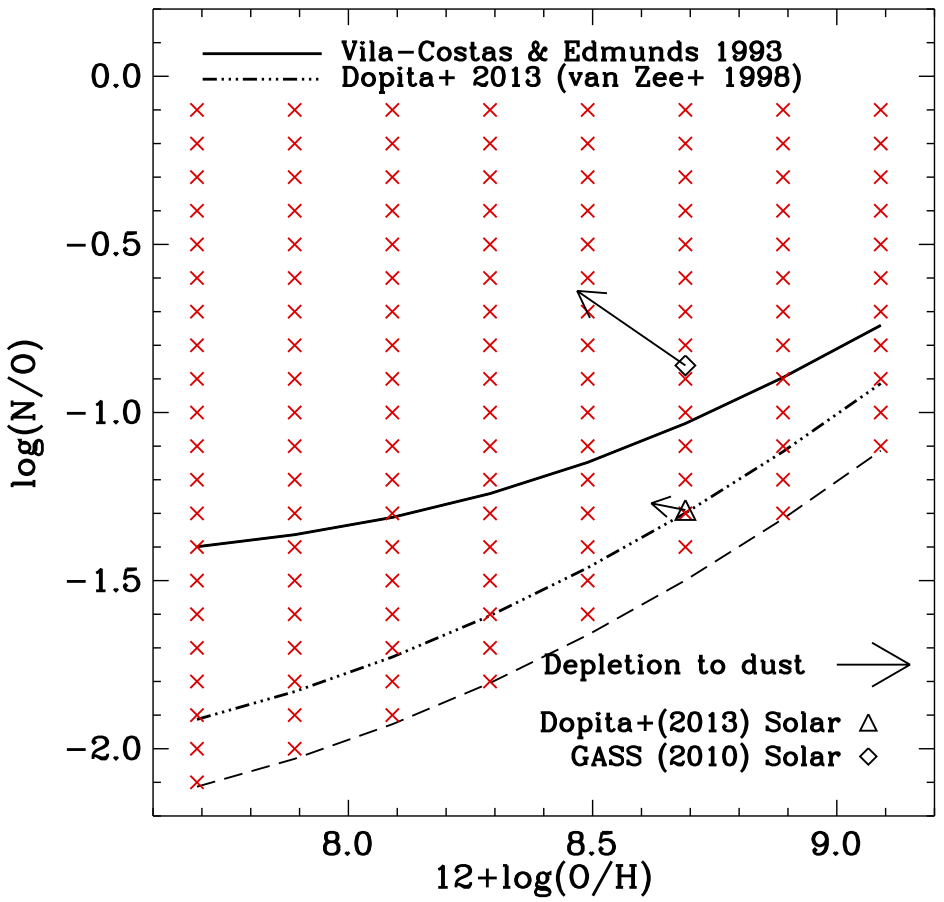}
\caption{The sampling of the N/O vs. O/H parameter space by our simulations. The solid curve indicates the relationship given by \protect\cite{VilaCostasE93}; the dot-dot-dot-dashed curve indicates the relationship used by \protect\cite{Dopita13} which is based on a fit to the measurements from \protect\cite{vanZee98b}. The dashed curve at the bottom indicates the lower limit we placed, which is 0.2 dex lower than the \protect\cite{Dopita13} relation. The red crosses indicate all the abundance patterns we simulate. The diamond symbol indicates the solar N/O and O/H values given by \protect\cite{GrevesseASS10}, and the triangle symbol indicates the solar values used by \protect\cite{Dopita13}. The latter reference used the \protect\cite{GrevesseASS10} solar values for all the other elements heavier than Helium, except for Nitrogen. The arrows indicate the change to the abundance pattern due to depletion onto dust. The abundance quoted in this figure and throughout this paper are the abundance before depletion onto dust.}
\label{fig:no_oh_space}
\end{center}
\end{figure}


Many theoretical modeling efforts, \citep[e.g.][]{KewleyD02, Dopita13} for star-forming galaxies have assumed a fixed relationship between N/O vs. O/H given empirical measurements and the expectation of the secondary nucleosyntheis contribution of Nitrogen. However, in galaxies with different chemical enrichment history, the resulting N/O vs. O/H could be very different from the typical trend. In quiescent galaxies, this could also be very different from what we see in star-forming galaxies as we do not yet know the origin of the warm ionized gas.

Here we do not limit our models to a fixed N/O vs. O/H relationships. Instead, we sample the N/O vs. O/H parameter space using 122 different combinations for these two parameters. They are illustrated in Fig.~\ref{fig:no_oh_space}. \cite{VilaCostasE93} and \cite{vanZee98b} both derived N/O vs. O/H for a large number of HII regions in nearby galaxies. However, the relationship they obtain are quite discrepant. The solid curve in Fig~\ref{fig:no_oh_space} shows the relationship derived by \cite{VilaCostasE93}, the dot-dot-dashed curve shows the fitting formula derived by \cite{Dopita13} based on the measurements from \cite{vanZee98b}. One can see that they differ substantially, up to 0.5 dex at the low metallicity end. \cite{VilaCostasE93} results were based on the formula from \cite{Pagel92} which are derived from old theoretical calculations. \cite{vanZee98b} results were bootstrapped from many empirical calibrations. Only a small set of \cite{vanZee98b} results were based on direct temperature measurements and an approximate relationship relating O$^{++}$ zone temperature with N$^+$ zone temperature. Most of their measurements were instead based solely on strong lines. In addition, neither efforts corrected for the effect of depletion of elements onto dust. Therefore, it is not surprising to find large discrepancies. 

In order to fully sample the potential parameter space, we set a lower limit by subtracting 0.2 dex in N/O from the \cite{Dopita13} relation. Then we sample all the N/O vs. O/H combinations up to $\log (N/O)=-0.1$. In O/H, we go from 0.1 to 2.5 times the solar O/H value. The combinations are indicated by the red crosses.  We also mark the solar abundance values from \cite{GrevesseASS10} and that used by \cite{Dopita13}. The latter used \cite{GrevesseASS10} abundance for all other elements heavier than Helium, except for Nitrogen.

We would also like to point out that the depletion of elements on dust grains can significantly impact the resulting line ratios. One should be clear on whether the derived gas-phase abundance includes the depleted portion or not. Here we adopt the default depletion factors in the Cloudy C17.00 code, where the Oxygen abundance is reduced by 40\% (-0.22 dex) and Nitrogen abundance is unchanged (not depleted). This is very different from those assumed by \cite{Dopita13} where Oxygen is depleted by 0.07 dex and Nitrogen by 0.05 dex. {\it The abundance quoted throughout this paper are the abundances before depletion onto dust.}

Following \cite{Dopita13}, we also change the Carbon abundance accordingly, based on observations by \cite{Garnett04}. We set the Carbon abundance to be 0.6dex higher than the Nitrogen abundance. For all the other elements heavier than Helium, the abundances are set to scale with Oxygen following the solar pattern given by \cite{GrevesseASS10}. Helium abundance is set to a constant according to the solar value. 

The input ionizing spectrum is assumed to be that of a simple stellar population that is 13 Gyr old. The spectrum is made using \cite{BC03} stellar evolution synthesis code with solar metallicity and Chabrier initial mass function. We vary the ionization parameter ($\log U$) from -4.5 to -2 with 0.5 dex spacing. 

The simulation includes photoelectric heating by dust and attenuation by dust inside the ionized regions. The type of grains are set to be representative of the Milky Way ISM. The number of grains are set to scale with the Oxygen abundance. The simulation is run under constant gas pressure with an initial Hydrogen density of 100 cm$^{-3}$, and assuming an open geometry. We include the galactic background cosmic rays and the cosmic radio to X-ray background  in the simulation, using the default values provided by Cloudy. 

We also run the simulation for a different initial density (n=10/cm$^3$) and a different input ionizing spectrum (3Myr old SSP) to check the reliability of our assumptions. 

\section{Fitting relations from the model grids}\label{sec:calibration}

\begin{figure}
\begin{center}
\includegraphics[width=0.48\textwidth]{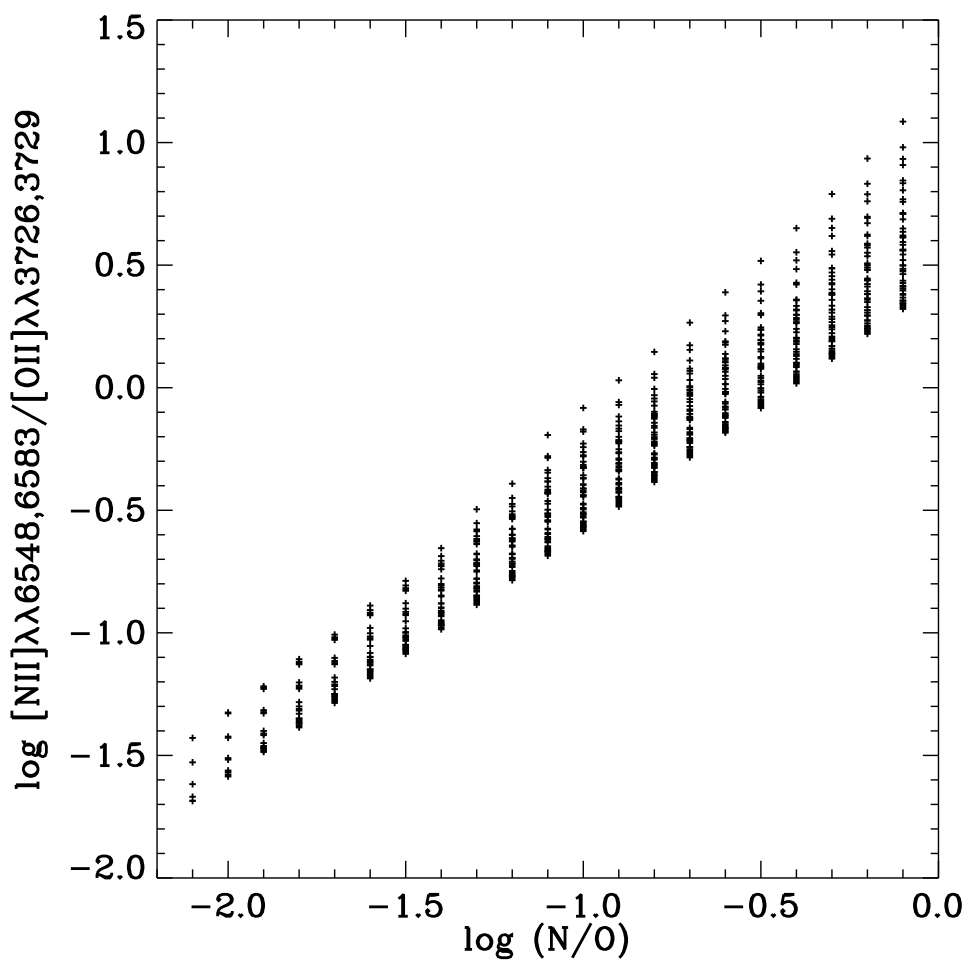}
\caption{\nii/\oii\ vs. N/O for our photoionization models. Although \nii/\oii\ is strongly correlated with N/O ratio, at each N/O ratio there can be a range of \nii/\oii\ ratio produced. The vertical scatter is primarily caused by temperature difference as we will demonstrate below. The temperature difference could come from ionization parameter difference and/or abundance pattern variation.}
\label{fig:n2o2_no_model}
\end{center}
\end{figure}

Figure~\ref{fig:n2o2_no_model} shows how \niiw/\oiiw\ respond to N/O ratio variations. There is a strong correlation but there are significant scattering, i.e., models with the same N/O ratio can span a range in \nii/\oii. This scatter is primarily caused by different temperatures among the models. This is because this line ratio is determined by the relative numbers of N$^+$ and O$^+$ ions and the temperatures inside the N$^+$ and O$^+$ zones. Because N and O have very similar ionization potentials, N$^+$ and O$^+$ are almost always co-spatial. Thus, N$^+$/O$^+$ trace N/O very closely. At fixed N/O ratio, N$^+$/O$^+$ are nearly the same as well. The difference in \nii/\oii\ is mostly driven by temperatures in the zone populated by N$^+$ and O$^+$. 


\begin{figure*}
\begin{center}
\includegraphics[width=0.9\textwidth]{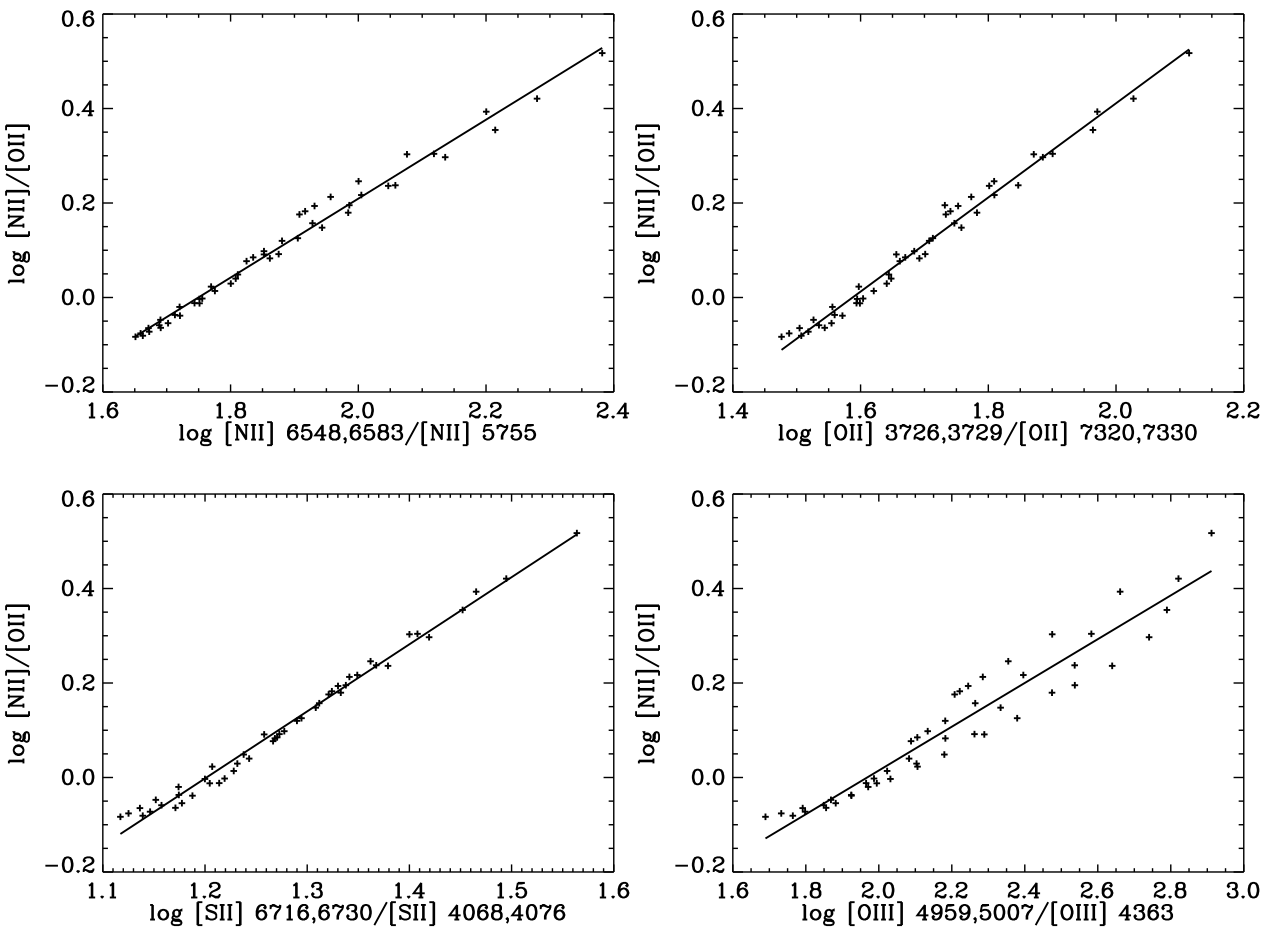}
\caption{At fixed N/O abundance ratio ($\log (N/O)=-0.5$ in this plot), \niiw/\oiiw\ ratio is well correlated with the temperature trace of N$^+$, O$^+$, and S$^+$, but poorly correlated with the temperature tracer of O$^{++}$. The relationship for other values of N/O ratio share the same slope with different intercepts.}
\label{fig:n2o2_temp_fixed_no}
\end{center}
\end{figure*}

We pick a group of models that have the same N/O ratio and study the dependence of \nii/\oii\ on temperature-sensitive line ratios.  Figure~\ref{fig:n2o2_temp_fixed_no} shows the scatter in \niiw/\oiiw, at fixed N/O abundance, is tightly correlated with the temperature-sensitive line ratios from S$^+$, N$^+$, or O$^+$, but poorly correlates with the temperature proxy for O$^{++}$. Therefore, it is much better to use the former in our derivation of the N/O abundance.

This also informs us that there is not a simple one-to-one relationship between the temperature in O$^+$ zone and the O$^{++}$ zone. Two models with the same O$^+$ zone temperature could have different O$^{++}$ zone temperatures, and vice versa, due to the different effects of ionization parameter variation and abundance pattern variation. Thus, inferring temperature of O$^+$ zone from temperature measurement for the O$^{++}$ zone could be dangerous, for both quiescent and star-forming galaxies. 


In practice, it is much easier to detect the \oiitw \footnote{Thoughout the paper, when we quote \oiitw, we include all 4 lines in the quadrulplet: 7318\AA, 7319\AA, 7329\AA, 7330\AA. The models and the measurements in the data also include any potential recombination contribution to these lines.} and \siitw\ lines than the \niitw\ line. The reason is that, for a fixed temperature, the line ratio contrasts between the strong nebular line and the weak auroral line for S$^+$ and O$^+$ are much smaller than that for N$^+$, making it easier to detect the weaker auroral line for S$^+$ and O$^+$. In fact, these are the only temperature-sensitive lines we reliably detected in stacked spectra of all subsamples of quiescent galaxies in SDSS. But there is a drawback for using these lines. The nebular line and the auroral line for S$^+$ or O$^+$ are widely separated in wavelength, making both temperature tracers very sensitive to dust extinction. This is probably one of the reasons that they are not often used for temperature measurements. 

We have a solution to this problem. Notice the relative wavelengths of the nebular line and auroral line for O$^+$ and S$^+$ are opposite to each other, making them having the opposite dependence on extinction. Therefore, it is possible to combine them to form a temperature tracer that is insensitive to extinction. Below, we use 7325 to denote the \oiitw\ which is actually a quadruplet, use 4072 to denote the \siitw, use 3727 to denote the \oiiw, and use 6720 to denote the \siiw. We use the subscript 0 to denote the intrinsic line ratio without dust extinction. 

\begin{eqnarray}
{I_{3727} \over I_{7325}} &= \left(\frac{I_{3727}}{I_{7325}}\right)_0 10^{-0.4(A_{3727}-A_{7325})} \\
{I_{6725} \over I_{4072}} &= \left(\frac{I_{6725}}{I_{4072}}\right)_0 10^{-0.4(A_{6720}-A_{4072})}
\end{eqnarray}

Here $A_\lambda$ denotes the extinction in magnitude at wavelength $\lambda$.  It is possible to cancel out the extinction term in a multiplication of these two line ratios. If we define
\begin{equation}
\alpha \equiv \frac{A_{3727}-A_{7325}}{A_{4072}-A_{6720}},
\end{equation}
we find that 
\begin{equation}
\left({I_{3727} \over I_{7325}}\right)\left({I_{6720} \over I_{4072}}\right)^\alpha = \left({I_{3727} \over I_{7325}}\right)_0\left({I_{6720} \over I_{4072}}\right)_0^\alpha.
\end{equation}

This combination of these two line ratios is independent of extinction. The parameter $\alpha$ has a very slight dependence on the assumed extinction curve. With the extinction curve given by \cite[CCM hereafter]{CardelliCM89}, we find $\alpha=1.297$ for $R(V)=3.1$ and $\alpha=1.307$ for $R(V)=5.0$. This variation of less than 1\% can be safely neglected given our expected precision of line measurements and abundance determination. We adopt $\alpha=1.30$ for the following analysis. We name this new composite line ratio ''SO$_{\rm T}$'' to indicate it is a temperature tracer based on sulfur and oxygen. 
\begin{equation}
{\rm SO}_{\rm T} \equiv \log \frac{I_{3727}}{I_{7325}} +1.3 \log \frac{I_{6720}}{I_{4072}}
\end{equation}.

\begin{figure}
\begin{center}
\includegraphics[width=0.48\textwidth]{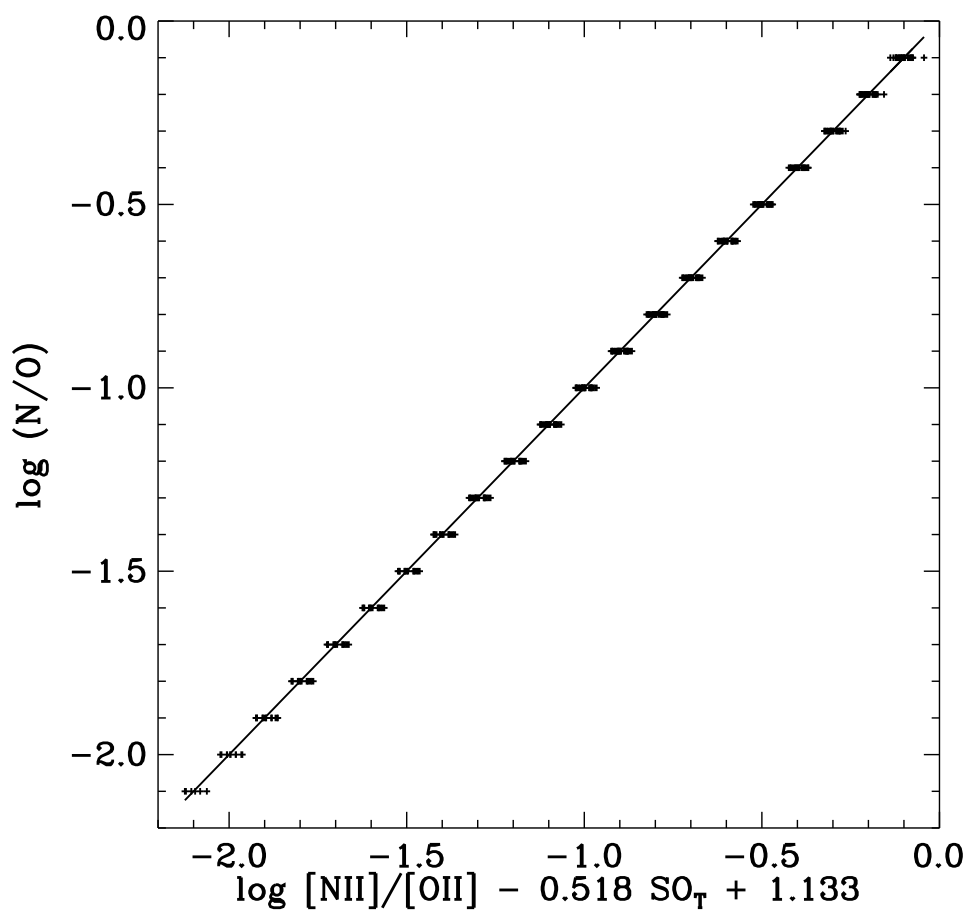}
\caption{The best fit relationship among N/O abundance ratio, \niiw/\oiiw, and the composite temperature tracer SO$_{\rm T}$. }
\label{fig:no_ct2}
\end{center}
\end{figure}

We fit the N/O ratio as a linear function of $\log \nii/\oii$ and SO$_{\rm T}$. The fit is shown in Fig.~\ref{fig:no_ct2}. The formula is
\begin{equation}
\log (\rm N/O) = \log \frac{\rm \niiw}{\rm \oiiw}-0.518 {\rm SO}_{\rm T} + 1.133  .
\label{eqn:no_sot_cal}
\end{equation}
Among the 792 models, the RMS residual of the fit is only 0.018 dex. 

Here we also provide the calibration based on \niiw/\niitw\ from our models. 
\begin{eqnarray}
\log (\rm N/O) =& 1.007\log \frac{\rm \niiw}{\rm \oiiw}\\ \nonumber
                &- 0.851 \frac{\niiw}{\niitw} + 0.992    .
\end{eqnarray}

We check whether our calibration depends on certain assumptions in our modelings. Using a similar simulation with slightly different density, n=10/cm$^3$, we obtained the following calibration
\begin{equation}
\log (\rm N/O) = 1.002\log \frac{\rm \niiw}{\rm \oiiw}-0.553 {\rm SO}_{\rm T} + 1.308 . 
\end{equation}

Compared to Eqn.~\ref{eqn:no_sot_cal}, the slopes derived are very similar; only the intercept has a 0.175 dex shift. The RMS residual around the best fit is 0.019 dex. 
It would introduce a systematic offset in our results but would not change the relative trend. 

We also checked whether changing the input ionizing spectrum to that of a star-forming region would affect the result significantly. Using a 3Myr SSP as input spectrum, we find the following relationship. 

\begin{equation}
\log (\rm N/O) = 1.017\log \frac{\rm \niiw}{\rm \oiiw}-0.601 {\rm SO}_{\rm T} + 1.560. 
\label{eqn:no_m3l_cal}
\end{equation}

Again, the slopes are similar. The intercepts has a shift of 0.5 dex. The residual RMS is 0.033dex. 
As we will see later, when applied to actual data, the differences in slope and intercept actually cancel out a bit, leading to smaller differences in absolute calibration. This demonstrates that our assumption is valid, that \nii/\oii\ combined with temperature tracer can probe N/O abundance regardless of the details of the ionization mechanism.

\section{Application to Quiescent Galaxies in SDSS}\label{sec:data}
Next, we apply our calibration to quiescent galaxies from the Sloan Digital Sky Survey (SDSS, \citealt{York00}). SDSS has imaged one quarter of the sky in 5 bands and took spectra for nearly a million galaxies and tens of thousands of quasars. We use the galaxy catalog from the New York University Value-added Galaxy Catalog \citep{BlantonSS05}. Using the same method as described in \cite{YanB12}, we have carefully measured several emission lines for each galaxy in the main sample of SDSS, after modeling and subtracting the stellar continuum. The measurements of these weak emission lines are sensitive to high-frequency flux calibration systematic errors and inaccuracies in the templates used to construct the stellar continuum. For the former, we have applied the correction vector derived by \cite{Yan11flux}. For the latter, we have statistically measured the systematic offset in the measurements of the emission line equivalent width and corrected the line flux accordingly \citep{Yan18a}. 

Quiescent galaxies usually have old stellar population which gives them red optical colors. Thus, to select them, we start with a sample of red galaxies that lie between redshift of 0.05 and 0.15.  This lower limit on redshift is chosen so that the 3\arcsec\ SDSS fiber aperture covers at least a diameter of 3 kpc (assuming a Hubble constant of 70km s$^{-1}$ Mpc$^{-1}$) so that the line emission is not dominated by the nuclear emission which may be powered by a different ionization mechanism. In the great majority of quiescent galaxies, the line emission is spatially extended and this extended component dominates over the nuclear component on scales greater than 1 kpc in diameter \citep{YanB12}. The higher limit in redshift is designed to ensure significant signal-to-noise in the emission line measurements.

We apply a color cut and a D4000 cut in the same way as done by \cite{Yan18a} to extract a very pure sample of quiescent galaxies. In this sample, 17.6\% of galaxies have \hal\ emission detected above $3\sigma$. All of these galaxies appear to fall in the LINER region of a BPT diagram, although they are not selected based on emission line ratios. This strongly suggests that they are powered by a common ionization mechanism. 

As a comparison, we select a sample of star-forming galaxies from SDSS. These are selected using a cut in BPT diagram. We require each galaxy to have fractional errors in both \niibw/\hal\ and \oiiibw/\hb\ smaller than 0.25 dex\footnote{We put the significance requirement on the error of the ratio, rather than on individual line detection significance, because the error of the ratio is more directly relevant to the results.}. Then we use the demarcation given by \cite{KauffmannHT03} to select only star-forming galaxies. 

\begin{figure*}
\begin{center}
\includegraphics[width=0.48\textwidth]{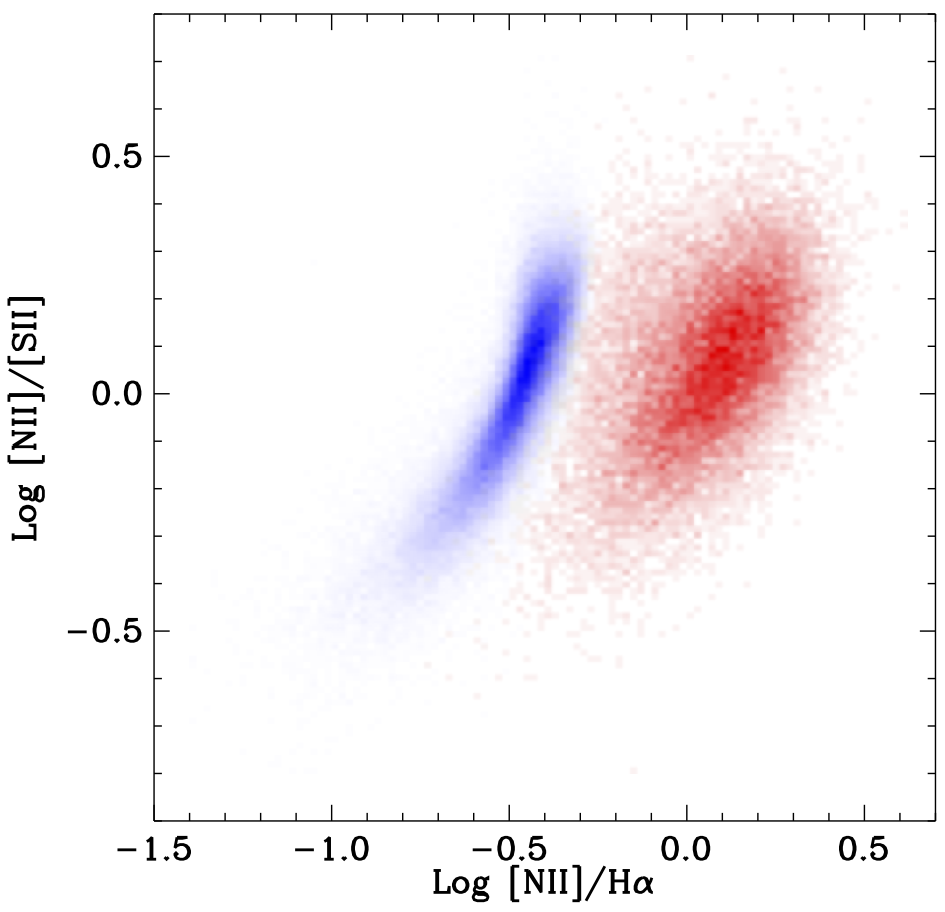}
\includegraphics[width=0.48\textwidth]{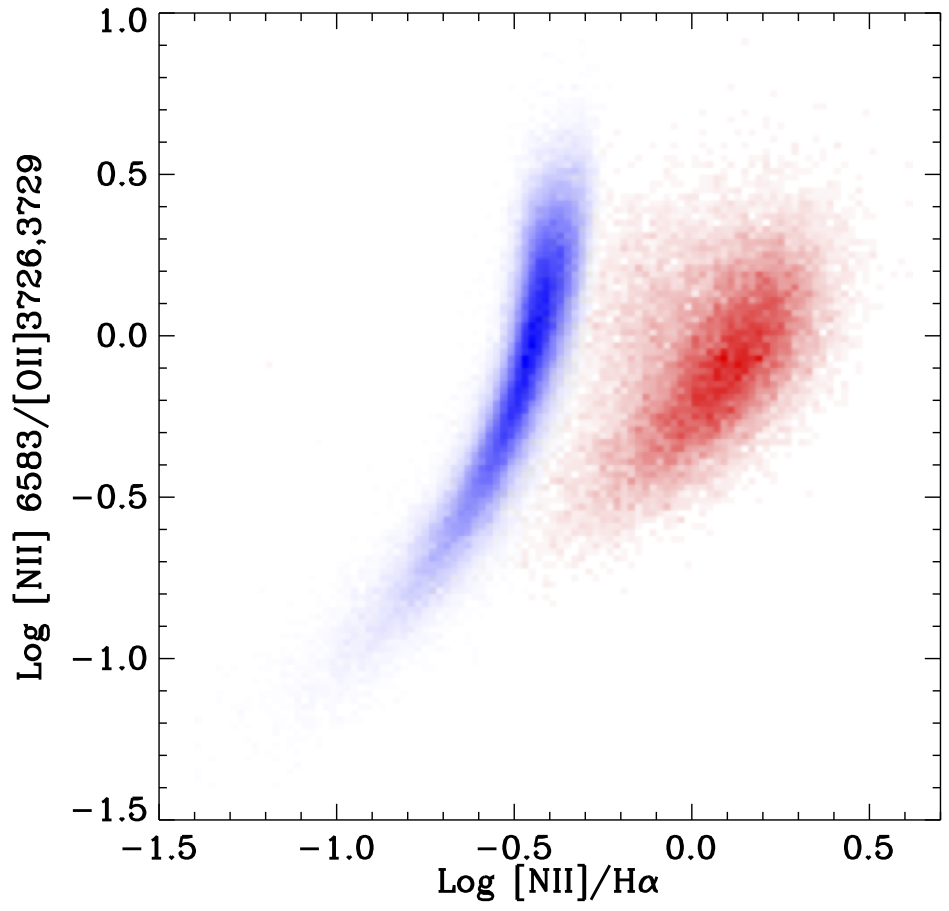}
\caption{Left: distribution of star-forming galaxies and quiescent galaxies in \niibw/\siiw\ vs. \niibw/\hal\ diagram. Right: distribution of star-forming galaxies and quiescent galaxies in \niibw/\oiiw\ vs. \niibw/\hal\ diagram. In both panels, the blue locus on the left is for star-forming galaxies and the red locus on the right is for quiescent galaxies. The loci of star-forming galaxies are primarily tracing O/H and N/O abundance variation. We argue the loci of quiescent galaxies are also tracing such abundance variation, at least in N/O ratio. }
\label{fig:n2ha_n2s2_n2o2}
\end{center}
\end{figure*}

In Fig.~\ref{fig:n2ha_n2s2_n2o2}, we show the \nii/\sii\ vs. \nii/\hal, and \nii/\oii\ vs. \nii/\hal\ for both the star-forming sample and the quiescent sample. Not all quiescent galaxies have significant emission lines to give us reliable line ratios. When selecting galaxies for these plots, we require the fractional error on both axes to be smaller than 0.25 dex. This selects 17.8\% of all quiescent galaxies.  The resulting quiescent sample has a median error of 0.15 dex in \niibw/\siiw\ and 0.14 dex in \niibw/\hal. Such selection could potentially bias the result. For this reason, we also checked the plots by selecting only those quiescent galaxies with \oii\ EW greater than 85-percentile among all quiescent galaxies. The resulting distribution are nearly the same. Thus, we conclude the selection based on uncertainty does not bias the results. 

In these two plots, the star-forming galaxies populate curved loci from the lower left to the upper middle. While the quiescent galaxies are populating the area to the right of the star-forming loci, with greater \nii/\hal\ ratios. The majority of the quiescent galaxies also display curved loci with a similar shape to that displayed by star-forming galaxies.  The star-forming loci are primarily tracing O/H and N/O variations in those galaxies. We strongly suspect the loci for the quiescent galaxies are also tracing abundance ratio variations. 

One may be worried that \nii/\oii\ can be strongly affected by dust extinction. Given the trend in \nii/\sii\ and \nii/\hal, both of which are insensitive to extinction, it is unlikely that most of the variation of \nii/\oii\ is due to extinction variation. In addition, quiescent galaxies span a similar range as star-forming galaxies do in \nii/\sii, but span a smaller range than star-forming galaxies in \nii/\oii. This strongly indicates that quiescent galaxies have much smaller extinction variation among them than that among star-forming galaxies. 

Below, we demonstrate that different parts of the quiescent locus indeed have different N/O ratios using the calibration derived in previous sections.





\begin{figure}
\begin{center}
\includegraphics[width=0.48\textwidth]{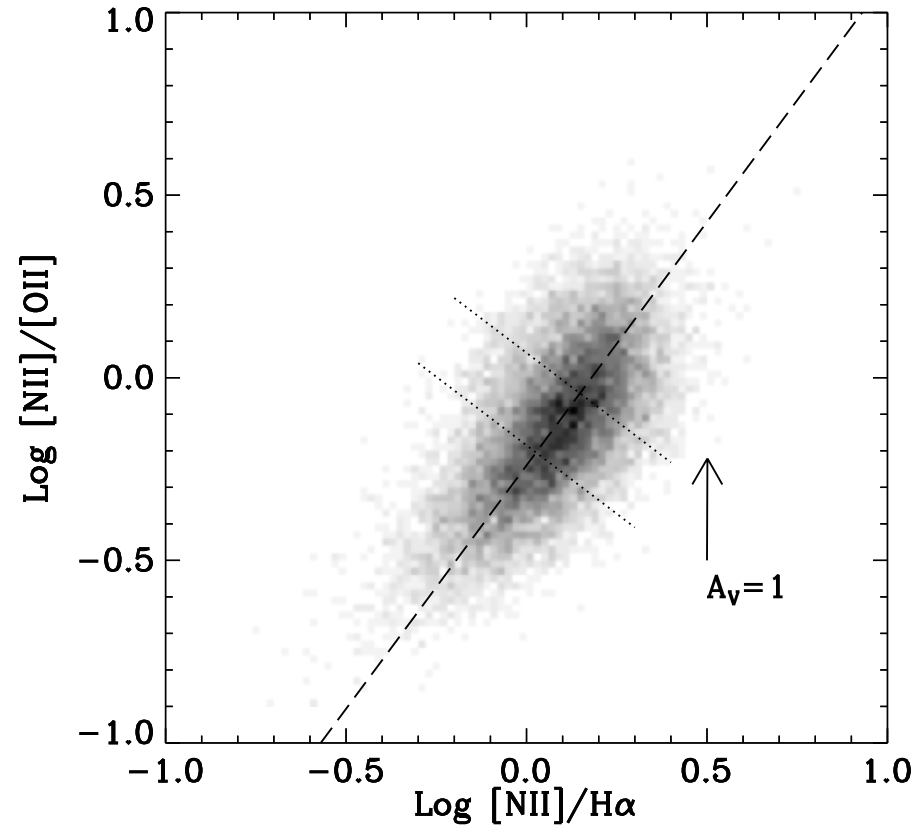}
\caption{Distribution of \nii/\oii\ vs. \nii/\hal\ of quiescent galaxies. Compared to quiescent galaxies in the right panel of Fig.~\ref{fig:n2ha_n2s2_n2o2}, the major difference is that we have applied an additional cut in \oii\ EW vs. \hal\ EW space to remove 3\% of the sample that populate the upper left region away from the main locus of the quiescent galaxies. The sample in this plot is the top 25\% in total emission line equivalent width. No cuts in fractional errors of \nii/\hal\ or \nii/\oii\ are applied. The two dotted lines indicate the thresholds used to split this sample into 3 subsamples with different line ratios.}
\label{fig:n2o2_n2ha_pure}
\end{center}
\end{figure}

In order to make reliable N/O ratio measurements, we need to measure temperature-sensitive auroral lines. In a companion paper, \cite{Yan18a}, we have measured the several weak auroral lines in the stacked spectra of quiescent galaxies. In that work, we selected a sample of quiescent red galaxies in the same way as described above, with an additional cut in \oii\ EW vs. \hal\ EW space to remove 3\% of the sample that include galaxies with Seyfert nuclei, galaxies with low level of star formation, very dusty galaxies, or galaxies with different abundance patterns. This results in a narrower sequence of galaxies in the \nii/\oii\ vs. \nii/\hal\ plot. We pick out galaxies that belong to the top 25\% in emission line equivalent width using a combined indicator from multiple lines. We show this sample in Fig.~\ref{fig:n2o2_n2ha_pure}. We split them into three strong-line subsamples along this sequence with high-, mid-, and low-\nii/\hal\ ratios. The cuts are shown in Fig.~\ref{fig:n2o2_n2ha_pure}. We then constructed corresponding subsamples with zero emission lines, matched to each of the strong-line subsample in absolute magnitude $M_{0.1r}$, $D_n(4000)$, and $V_{\rm disp}$. We stacked the spectra of all galaxies in each strong-line and zero-line subsample. Then we subtract the zero-line stack spectrum from its corresponding strong-line stack spectrum to arrive at emission-line only spectra for the three strong-line subsamples. We measured the strong and weak lines in each spectrum. The results are tabulated by \cite{Yan18a}. 

Even in the super high S/N stacked spectra, some of the auroral lines are still very challenging to detect. For \oiiitw, we only detected it above 3$\sigma$ in one of the subsamples. For \niitw, we only detected it above 3$\sigma$ in two of the subsamples. Fortunately, the \oiitw\ and \siitw\ are relatively stronger and they are significantly detected in all three stacked emission-line spectra. Therefore, we could apply the calibration derived in \S\ref{sec:calibration} to these subsamples to measure the average N/O abundance ratio for them.

\begin{table*}
\begin{tabular}{l|c|c|c}
\hline\hline
Lines & Hi-\nii/\hal\ & Mid-\nii/\hal\ & Low-\nii/\hal\  \\ \hline
\hal\ & $ 439.2\pm  8.0$& $ 393.6\pm  6.1$& $ 390.6\pm  5.7$\\
\hb\ &  $100$ & $100$ & $100$\\
\niibw\ & $ 687.7\pm 12.3$& $ 459.4\pm  7.0$& $ 312.1\pm  4.6$\\
\siiw\ & $ 466.9\pm  8.5$& $ 394.8\pm  6.2$& $ 363.1\pm  5.4$\\
\oiiw\ & $ 583.6\pm 10.6$& $ 620.9\pm  9.5$& $ 686.0\pm  9.8$\\
\siitw\ & $  24.3\pm  3.6$& $  25.4\pm  2.4$& $  19.7\pm  2.1$\\
\oiitw+\caiibw\ & $  24.6\pm  2.8$& $  23.2\pm  2.3$& $  18.9\pm  2.1$\\
\oiitw\ (corrected for \caiibw)& $24.6\pm  4.0$& $20.6\pm  2.5$& $  14.0\pm  4.5$\\\hline
\hb\ Flux &  $ 0.276\pm0.005$& $ 0.340\pm0.005$& $ 0.372\pm0.005$\\\hline\hline
\end{tabular}
\caption{Raw measurements of relevant emission lines in the continuum-subtracted stacked spectra from \protect\cite{Yan18a}, without any correction for extinction. The measurements are given relative to the flux of the \hb\ line, which is given in the bottom row. The unit of the \hb\ line fluxes is Angstrom multiplied with the flux density between 6000\AA\ and 6100\AA, since each spectrum is normalized by the median flux in the window 6000-6100\AA\ before being stacked. The \oiitw\ lines are blended with the \caiibw\ line, whose strength can be derived from the measurement of \caiiaw. We show the results of \oiitw\ both before and after correcting for the \caiibw\ contamination.}
\label{tab:lines}
\end{table*}

We reproduce some of the measured line strengths relevant for our goal here in Table.~\ref{tab:lines}.

\begin{figure}
\begin{center}
\includegraphics[width=0.49\textwidth]{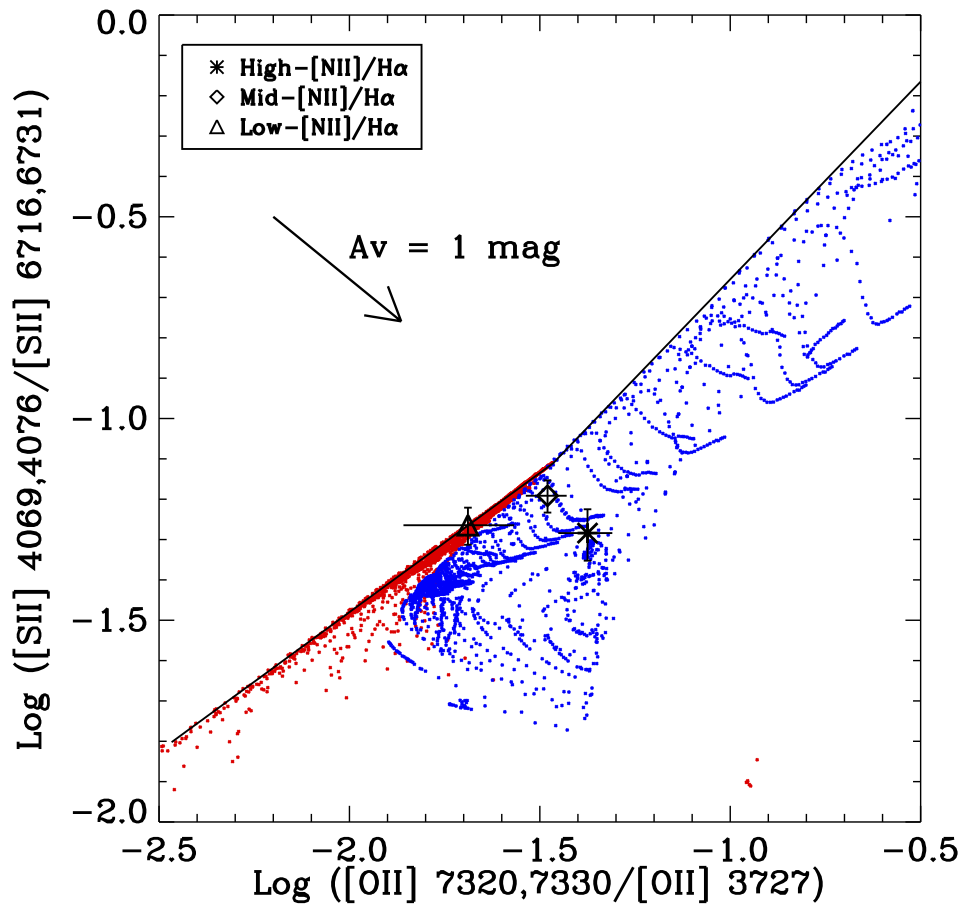}
\caption{Distribution of various models in \siitw/\siiw\ vs. \oiitw/\oiiw\ space. The red points show photoionization models described in this paper; the blue points show shock models from \protect\cite{Allen08}. The lines indicate approximate upper boundaries of these models. The arrow indicate the effect of dust extinction with $A_V=1$ mag. The data points with error bars are the measurements made by \protect\cite{Yan18a}.} 
\label{fig:st2_ot2_models}
\end{center}
\end{figure}

We first need to correct our \niiw/\oiiw\ ratios for extinction. Using the Balmer decrement and the CCM extinction curve, we found the $A_{\rm V}$ to be 1.33, 0.99, and 0.97 mag for the hi-, mid-, and low-\nii/\hal\ subsamples, respectively. However, we think these corrections are significantly overestimated. As shown by \cite{Dopita82}, the combination of \siitw/\siiw\ and \oiitw/\oiiw\ can also be used for dust extinction estimates. Here we develop this method further. In Figure~\ref{fig:st2_ot2_models}, we show the \siitw/\siiw\ vs. \oiitw/\oiiw\ for all the models we have run for different density, abundances, and input ionizing spectra. We also add in shock models from \cite{Allen08}.  We can see that a lot of the models populate the diagonal line from bottom left to upper right. Departures from this line can only happen towards the lower right. This is because the O$^+$ zone is always equal to or hotter than the S$^+$ zone, regardless of the ionization mechanism, as the ionization potentials for O$^0$ and O$^+$ are higher than those for S$^0$ and S$^+$. The upper left region of this diagram would be unphysical. Given these models, we found the following relationships for the approximate upper limit on \siitw/\siiw\ as a function of \oiitw/\oiiw, in the low density regime (around or below $100 {\rm cm}^{-3}$). 
\[
     \log {I_{4072} \over I_{6720}}= 
\begin{cases}
0.69 \log \frac{I_{7325}}{I_{3727}} - 0.1,& \text{if SO}_{\rm T} > 2.91  \\
0.98 \log \frac{I_{7325}}{I_{3727}} +0.325,& \text{otherwise}
\end{cases}
\]

Dust extinction would also shift points towards the lower right. Thus, we can use the upper boundary formed by the models in this plot to put an upper limit on dust extinction: 
\[
A_{\rm V} \leq 
\begin{cases}
1.408\log \frac{I_{7325}}{I_{3727}} -2.040 \log \frac{I_{4072}}{I_{6720}}-0.204,& \text{if SO}_{\rm T} > 2.91 \\
1.668\log \frac{I_{7325}}{I_{3727}} -1.702 \log \frac{I_{4072}}{I_{6720}}+0.553,& \text{otherwise}
\end{cases}
\]

Given the measurements in Table~\ref{tab:lines}, we find $A_{\rm V}$ upper limits of 0.48, 0.14, 0.0 mag for the high-, mid-, and low-\nii/\hal\ subsamples, respectively. The extinction would be of these values if the gas is photoionized, as almost all the photoionized models fall along the diagonal relation.  But these values would be upper limits if the gas is ionized by shocks. 
If we use these extinction upper limits to correct the \nii/\oii\ ratios, and apply Equation \ref{eqn:no_sot_cal}, we obtain their N/O abundances listed in Table.~\ref{tab:n_to_o}.  We found 0.35 dex of N/O abundance variation from the high-\nii/\hal\ to the low-\nii/\hal\ subsample. This should roughly represent the spread between 17-percentile and 83-percentile of the whole distribution. Due to the concentration of most galaxies in the middle range of \nii/\oii, the average N/O ratios of the subsamples are also quite concentrated to the middle. The tails of the distribution could have much larger or much smaller N/O ratios. 

\begin{table}
\begin{tabular}{l|c}
\hline\hline
Subsample & $\log{\rm (N/O)}$ \\ \hline
Hi-\nii/\hal &   $-0.39\pm0.06$ \\
Mid-\nii/\hal &   $-0.46\pm0.04$ \\ 
Low-\nii/\hal &   $-0.74\pm0.06$ \\
\hline\hline
\end{tabular}
\caption{N/O abundance measurements for the three subsamples.}
\label{tab:n_to_o}
\end{table}

Note the absolute values should be taken with a grain of salt given the uncertainty of the ionization mechanism, but the relative differences should be robust. If instead, we use the calibration derived from a different simulation (Eqn.~\ref{eqn:no_m3l_cal}, the absolute values would be higher by 0.18 dex for the hi- and mid-\nii/\hal\ subsamples, and by 0.15 dex for the low-\nii/\hal\ subsample.

One may wonder why the extinction corrections derived from coronal-to-strong line ratios of [SII] and [OII] are much smaller than those derived from \hal/\hb\ ratios. One possibility is that there may still be stellar absorption residuals under the Balmer lines in the emission-line only spectra, affecting just the Balmer lines but not the other forbidden lines. Another possibility is that the intrinsic \hal/\hb\ ratio may be significantly enhanced due to collisional excitation.

\section{Summary}
Gas phase abundance patterns in quiescent galaxies are difficult to explore in the optical wavelengths due to the uncertainty of the ionization mechanism. Despite this uncertainty, we believe one can make progress on some abundance ratios, such as N/O, based on simple physical arguments that apply to multiple ionization mechanisms. In this paper, we have developed a simple N/O abundance ratio calibration for quiescent galaxies. It is a direct T$_e$-based method as it requires the detection of temperature-sensitive auroral lines. We have devised a new extinction-insensitive temperature tracer based on auroral lines from O$^+$ and S$^+$. We also explored a wide range of parameter space in N/O vs. O/H to prove that a reliable measurement of N/O can be made independent of the assumed N/O vs. O/H relationship. We have applied our calibration to quiescent galaxies in SDSS and found they have a 0.35 dex spread in N/O ratio between 17- and 83-percentiles. This will help shed light on the chemical enrichment history of the warm ionized gas in these galaxies. 

One may want to take this method further and measure O/H abundance in these quiescent galaxies. We refrain from doing this because that derivation would depend strongly on the assumption of the ionization mechanism. We warn the reader from taking this method too far before confirming the ionization mechanism. 



\section*{Acknowledgements}
The author acknowledges the support of NSF Grant AST-1715898. 

    Funding for the SDSS and SDSS-II has been provided by the Alfred P. Sloan Foundation, the Participating Institutions, the National Science Foundation, the U.S. Department of Energy, the National Aeronautics and Space Administration, the Japanese Monbukagakusho, the Max Planck Society, and the Higher Education Funding Council for England. The SDSS Web Site is http://www.sdss.org/.

    The SDSS is managed by the Astrophysical Research Consortium for the Participating Institutions. The Participating Institutions are the American Museum of Natural History, Astrophysical Institute Potsdam, University of Basel, University of Cambridge, Case Western Reserve University, University of Chicago, Drexel University, Fermilab, the Institute for Advanced Study, the Japan Participation Group, Johns Hopkins University, the Joint Institute for Nuclear Astrophysics, the Kavli Institute for Particle Astrophysics and Cosmology, the Korean Scientist Group, the Chinese Academy of Sciences (LAMOST), Los Alamos National Laboratory, the Max-Planck-Institute for Astronomy (MPIA), the Max-Planck-Institute for Astrophysics (MPA), New Mexico State University, Ohio State University, University of Pittsburgh, University of Portsmouth, Princeton University, the United States Naval Observatory, and the University of Washington.

\bibliographystyle{mnras}
\bibliography{astro_refs}

\bsp
\label{lastpage}

\end{document}